\documentclass{article}

\usepackage [utf8]{inputenc}
\usepackage [T1] {fontenc}
\usepackage{xspace} 
\usepackage{graphicx} 
\usepackage[linktocpage]{hyperref} 
\usepackage{enumerate}

\usepackage[toc,page]{appendix} 

\usepackage{geometry}
\usepackage{caption} 
\usepackage{subcaption}

\newcommand{\R}{\mathbb{R}}

\newcommand{\N}{\mathbb{N}}
\newcommand{\trans}{^{\top}}

\newcommand{\bs}{\boldsymbol}

\title{Trans-Gaussian Kriging in a Bayesian framework : a case study}
\author{Joseph Muré}

\usepackage{amsthm}
\usepackage{amsmath}
\usepackage{amssymb}

\usepackage{natbib}
\newtheorem{thm}{Theorem}

\newtheorem{prop}[thm]{Proposition}

\newtheorem{defn}[thm]{Definition}

\theoremstyle{remark}

\let\lvert=|\let\rvert=|

\newcommand{\corr}{ \bs{ \Sigma }_{ \bs{ \theta } } }

\begin{document}

\maketitle

\begin{abstract}
In the context of Gaussian Process Regression or Kriging, we propose a full-Bayesian solution to deal with hyperparameters of the covariance function. This solution can be extended to the Trans-Gaussian Kriging framework, which makes it possible to deal with spatial data sets that violate assumptions required for Kriging. It is shown to be both elegant and efficient. 
We propose an application to computer experiments in the field of nuclear safety, where it is necessary to model non-destructive testing procedures based on eddy currents to detect possible wear in steam generator tubes.
\end{abstract}

\section{Introduction}

Non-destructive testing (NDT) is a group of techniques used in industry to evaluate the properties of components or systems without destroying them. Numerical simulations can be used to calibrate NDT techniques and determine adequate threshold levels for defect detection. These simulations can be computationally expensive. As a result, we may want to replace them by a cheaper surrogate model. \medskip

The NDT problem that motivates this paper is the following one.
In order to improve nuclear safety, Électricité de France (EDF) has developed a Non-Destructive Evaluation for testing the presence of a defect in Steam Generator Tubes.
In nuclear power plants, steam generators are the interface between primary and secondary water circuits. High-pressured water from the primary circuit flows into the steam generator tubes. 

To prevent the tubes from vibrating under this solicitation, they are held in place by anti-vibration bars (AVB). Rubbing may in time leave defects in the tubes, however. To detect them, a SAX (axial) probe is moved down the tube. This Non-Destructive Exam was modeled in C3D, which can accurately represent any possible defect geometry \citep{MCT13}. As this code uses the finite element method, a wide range of parameters may be taken into account. Furthermore, any degree of accuracy can be reached as long as the mesh is fine enough \citep{TGMC15}.
Non-destructive inspections based on eddy currents exploit the way the induction flux changes as the probe approaches a defect. If the tube were a perfect cylinder, the coils of the probe would get the same flux by cylindrical symmetry. A large enough difference between both fluxes therefore signals a defect. This differential flux is a complex quantity, but in the presented application only its imaginary part is used for defect detection, or rather 

the difference between maximum and minimum imaginary part 
as the probe moves through the tube. It is then converted to a tension by Lenz's law. \medskip

\begin{figure}[!ht]
\begin{center}

\includegraphics[angle=0,scale=0.5]{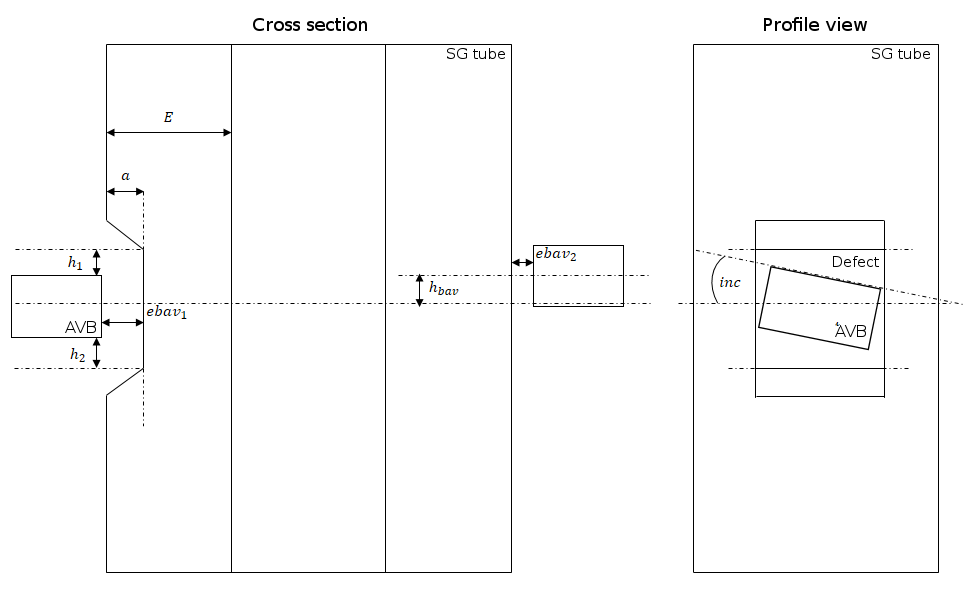}
\caption{Geometrical parameters of the computer model C3D. Left: Profile view of a steam generator tube where the left AVB caused wear. Right: View of the side with the defect.}
\label{Fig:test_illustration}
\end{center}
\end{figure}

Following both expert reports and data simulations, eight geometrical parameters (see Figure \ref{Fig:test_illustration}) and one non-geometrical parameter were identified as having the greatest influence on the output of C3D. In order to be able to define POD curves, they were given probability distributions reflecting expert opinion. 

\begin{enumerate}
\item $a \sim U[a_{a} , b_{a} ]$: defect depth (mm);
\item $E \sim N (a_e , b_e )$: pipe thickness (mm) based on measurements of 5000 pipes;
\item $h_{1} \sim U[a_{h_{1}} , b_{h_{1}} ]$: distance between the AVB and the top
of the defect (mm);
\item $h_{2} \sim U[a_{h_{2}} , b_{h_{2}} ]$: distance between the AVB and the
bottom of the defect (mm);
\item $e_{BAV1} \sim  U[-a + a_{e_{BAV1}} , b_{e_{BAV1}} ]$: length of the gap
between the AVB on the side of the defect and the tube (mm);
\item $e_{BAV2} \sim U[a_{e_{BAV2}} , b_{e_{BAV2}} ]$: length of the gap
between the AVB on the other side and the tube (mm);
\item $h_{BAV} \sim  U[a_{h_{BAV}} , b_{H_{BAV}}]$: shift between both AVBs (mm);
\item $inc \sim \mathcal{N}_{trunc,0}(inc_a, inc_b)$: inclination of the AVB on the side of the defect. $\mathcal{N}_{trunc,0}(inc_a, inc_b)$ denotes a Normal distribution with mean $inc_a$ and variance $inc_b$ truncated at 0: its support is $[0,+\infty)$ (mm);
\item $cond \sim \mathcal{N}(cond_a,cond_b)$: conductivity of the tube (MS/m).
\end{enumerate}

The parameter of interest being $a$, all other parameters are collectively denoted by the eight-dimensional vector $\bs{x} = (E, h_{1},h_{2},e_{BAV1},e_{BAV2},h_{BAV},inc,cond ) \trans$. $\bs{X}$ is the corresponding random variable: its joint distribution is the product of the distributions described above. For ease of manipulation, $\bs{X}$ is reparametrized in such a way as to follow the Uniform distribution on the 8-dimensional cube $(0,1)^8$. We also reparametrize a in the same way. The whole input space becomes the unit cube $(0,1)^r$ with $r=9$.  \medskip

Our ultimate goal is to compute the Probability Of Detection (POD) of a defect as a function of its depth $a$. We formalize this notion in Section \ref{Sec:POD}. Practical computation of the POD requires the use of surrogate models. Section \ref{Sec:Objective_Bayes_Trans_Gaussian_Kriging} provides the mathematical framework for dealing with surrogate model uncertainty. Finally, Section \ref{Sec:Industrial_application} presents the resulting POD curves in the case of steam generator defect detection.

\section{Probability Of Detection (POD)} \label{Sec:POD}

Industrial practice often uses Functional Risk Curves (FRC) as a means of expressing the ``probability'' of some (un)desirable event based on the value of some critical parameter. Probability Of Detection (POD) curves are a particular kind of FRC. They  represent the probability that a given testing procedure has of detecting a defect as a function of some parameter of interest. Two factors can justify the probabilistic framework:

\begin{enumerate}
\item the testing procedure may incorporate some randomness;
\item the result may depend not only on the parameter of interest but also on unknown nuisance parameters.
\end{enumerate}

We denote by $a$ the parameter of interest and by $\bs{x}$ the nuisance parameters. The set of all values possibly taken by $\bs{x}$ is endowed with a probability distribution which should reflect their frequency in real life. Let $\bs{X}$ be a random variable following this probability distribution.

For given $a$ and $\bs{x}$, $z(a,\bs{x})$ denotes the output of the testing procedure. 
 
Depending on the specific testing procedure used, it may or may not be random. In any case, it falls to the modeler to specify the probability distribution of $\bs{X}$ and/or the probability distributions of $z(a,\bs{x})$ for all $a$ and $\bs{x}$. In this paper, we only deal with deterministic testing procedures so POD curves are entirely determined by the distribution of $\bs{X}$. \medskip

Typically, the output $z$ is a scalar quantity. When this quantity lies beyond a predetermined threshold $s$, it is taken to denote the presence of a defect, so the POD curve is the function defined by

\begin{equation}
POD(a) = \mathbb{P}(z(a,\bs{X}) > s).
\end{equation}

Being a deterministic mapping does not make $z$ easy to determine, however. Computing the POD curve would in theory require conducting physical tests on a wide sample of objects with a wide variety of defects -- i.e. a wide variety of parameters $a$ and $\bs{x}$. Because of the prohibitive costs involved, phenomenological numerical simulations are used to simulate the testing procedure. Throughout this paper, such simulations -- collectively called a ``computer experiment'' -- are assumed to be perfectly accurate. \medskip

Unfortunately, even these numerical simulations are too costly to allow a Monte-Carlo approach. Therefore, $\bs{z}$ needs to be approximated by some less costly model. In order to control the ensuing error, statistical surrogate models are used -- linear regression or polynomial chaos for example. In this paper, we focus on Kriging, otherwise known as Gaussian Process regression, and on Trans-Gaussian Kriging, where a transformation step is performed on the output space before Kriging is performed. When a surrogate model is used, $z$ must be replaced by $Z$, a random mapping whose probability distribution represents the uncertainty about $z$. \medskip

Surrogate models introduce some ambiguity in the definition of the POD curve. Should it be approximated by $POD(a) \approx \mathbb{P}(Z(a,\bs{X}) > s)$ ? Although natural, this approximation conflates the uncertainty about $\bs{X}$ and about $Z$, although the two are different in nature. The distribution of $\bs{X}$ is understood in \textit{frequentist} terms: over the year, an experimenter conducting tests on multiple pieces of equipment will encounter varied values of $\bs{X}$. The distribution of $Z$ represents \textit{epistemic} uncertainty, which is irrelevant to the actual tests conducted in real life and could be mitigated if more computational resources were available. This point is illustrated in the description of a concrete application below. \medskip

Kriging and Trans-Gaussian Kriging surrogate models depend on parameters, which will henceforth be named ``hyperparameters'' to differentiate them from the ``physical'' parameters $a$ and $\bs{X}$. Hyperparameters are tricky, because they have tremendous influence so careful surrogate model calibration is normally required. The Bayesian paradigm, being a complete inferential approach (\citet{RMR11} provides a review of its range from a practical standpoint) can be used to avoid this step. In this paper, we propose eliminating hyperparameters from the model by means of Bayesian averaging out. Prior elicitation is done with the help of Bernardo's reference prior theory \citep{Ber05}.

\section{An Objective Bayesian outlook to Trans-Gaussian Kriging} \label{Sec:Objective_Bayes_Trans_Gaussian_Kriging}

\subsection{Kriging Likelihood} \label{Sec:Kriging_likeilhood}

For the sake of simplicity, we assume that the parameter of interest $a$ belongs to $[0,1]$ and that $\bs{x}$, which represents all nuisance parameters belongs to $[0,1]^{r-1}$ for some positive integer $r$. The scalar output of interest is $y(a,\bs{x})$.

Kriging is a very flexible surrogate model for computer experiments \citep{SWN}.
To use it, the computer experiment must first be conducted for $n \in \N$ values $(a^{(i)},\bs{x}^{(i)})$ (called observations points) of the parameters. The set of observation points $\left((a^{(i)},\bs{x}^{(i)}) \right)_{i \in [\![1,n]\!]}$ is called the design set. The vector of outputs $\bs{y} = (y(a^{(1)},\bs{x}^{(1)}),...,y(a^{(n)},\bs{x}^{(n)})) \trans \in \R^n$ is called the observation vector. \medskip

Kriging models the uncertainty about $y$ by defining $Y$ as a Gaussian process subjected to the condition that for every integer $i \in [\![1,n]\!]$, $Y(a,\bs{x}_i)=y(a,\bs{x}_i)$. Because this conditioning is critical, we need a way to differentiate the conditioned from the unconditioned process, so let $\mathcal{Y}$ be the unconditioned Gaussian process.

In the literature, $\mathcal{Y}$ is often assumed to be stationary. That is, the distribution of the unconditioned Gaussian Process (before knowing the observed values) should be invariant by translation. This simplifying assumption is often reasonable, because should the emulated function be non-stationary, 
the distribution of the Gaussian Process conditional to the observations would reflect this non-stationarity, provided the number of observations is sufficient. Moreover, assuming non-stationarity would require some sort of prior knowledge of the kind of non-stationarity that is expected. For example, warped Gaussian processes \citep{SGR04} use the knowledge of the presence of a discontinuity. \medskip

In some contexts, an additional assumption can be made: that the distribution of the Gaussian Process before observing the data is isotropic. When it is possible to make this assumption, the correlation kernel of the Gaussian Process is usually parametrized by two positive hyperparameters: variance $\sigma^2$ and correlation length $\theta$. Seeing the Gaussian Process as a response surface, one may think of $\sigma := \sqrt{\sigma^2}$ as the scale of variation of the output and of $\theta$ as the scale of variation of the input. \medskip

While isotropy is a natural assumption in geostatistics -- the original application of Kriging \citep{Mat60} --  it is rarely adequate when dealing with computer experiments. Each of the $r$ dimensions in the input space $[0,1]^r$ corresponds to one parameter, and the parameters can be heterogeneous. In such contexts, a correlation length $\theta_i$ is necessary for every parameter ($i \in [\![1,r]\!]$). The covariance kernel is then anisotropic and we denote by $\bs{\theta}$ the vector $(\theta_i)_{i=1}^r$. The best kind of anisotropic kernel for interpretation is anisotropic geometric, but tensorized kernels are often used for simplicity \citep{Ste99}[page 54]. \medskip

The easiest way to introduce non-stationarity is through the addition of some non-constant deterministic function. 
 In the Universal Kriging framework, this function is assumed to belong to a given small-dimensional vectorial space $\mathcal{F}_p$. One assumes therefore that there exists in $\mathcal{F}_p$ a function $f$ that adequately approximates the deterministic function one seeks to emulate, and the stationary Gaussian process then merely models our perception of the error made when using such an approximation. \medskip 

Let $(f_j)_{j=1}^p$ be a basis of $\mathcal{F}_p$ and let $\bs{\beta}=(\beta_j)_{j=1}^p \in \R^p$ be the vector such that
$
f = \sum_{j=1}^p \beta_j f_j.
$

Naturally, $p$ should be smaller than the number of observation points $n$ or the model would not be identifiable. Denote by $\bs{H}$ the $n \times p$ matrix whose $(i,j)$-th element is $f_j \left( \bs{x}^{(i)} \right)$. Again for the sake of identifiability, assume that $\bs{H}$ is of full rank. Let then $\bs{y}$ be the vector of length $n$ whose $i$-th element is the observation of the Gaussian process made at $\bs{x}^{(i)}$. It is a Gaussian vector with mean vector $\bs{H} \bs{\beta}$. Let $\corr$ be its correlation matrix.

The likelihood of the parameters $\bs{\beta}$, $\sigma^2$ and $\bs{\theta}$ when observing $\bs{y}$ is then:

\begin{equation} \label{Eq:vraisemblance}
L( \bs{y} \; | \;  \bs{\beta}, \sigma^2 , \bs{ \theta } ) = 
\left( \frac{ 1 }{ 2 \pi \sigma^2 } \right) ^ { \frac{n}{2} } | \bs{ \Sigma }_{ \bs{ \theta } } | ^ {- \frac{1}{2} } \exp \left\{ - \frac{ 1 }{ 2 \sigma^2 } \left(\bs{y} - \bs{H \beta} \right) \trans \bs{ \Sigma }_{ \bs{ \theta } }^{-1} \left( \bs{y} - \bs{H \beta} \right) \right\} \; .
\end{equation}

\subsection{Transformation}

If even more flexibility is needed, like in some cases arising naturally from examples in the natural sciences -- see the example below -- one can relax the stationarity assumption and even the assumption that the random process is Gaussian through Trans-Gaussian Kriging. The idea is to assume that a random field $\mathcal{Z}(a,\bs{x})$ \textit{would be} Gaussian and stationary if the output space $O$ (which is assumed to be an interval of $\R$) were transformed in an adequate fashion. Practically speaking, one must choose a parametric family of nondecreasing differentiable transformations  $g_\alpha: O \rightarrow \R $. 
For some $\alpha$, $\mathcal{Y} := g_\alpha(\mathcal{Z})$ is a stationary Gaussian random field. \medskip

The detection of defects on Steam Generator tubes is most importantly impacted by the depth of the defect. As a first approximation, one may say that the greater the defect depth, the greater the chances of detecting it. \medskip

\begin{figure}[!ht]
\includegraphics[angle=0,width=0.33\linewidth, height=0.33\linewidth]{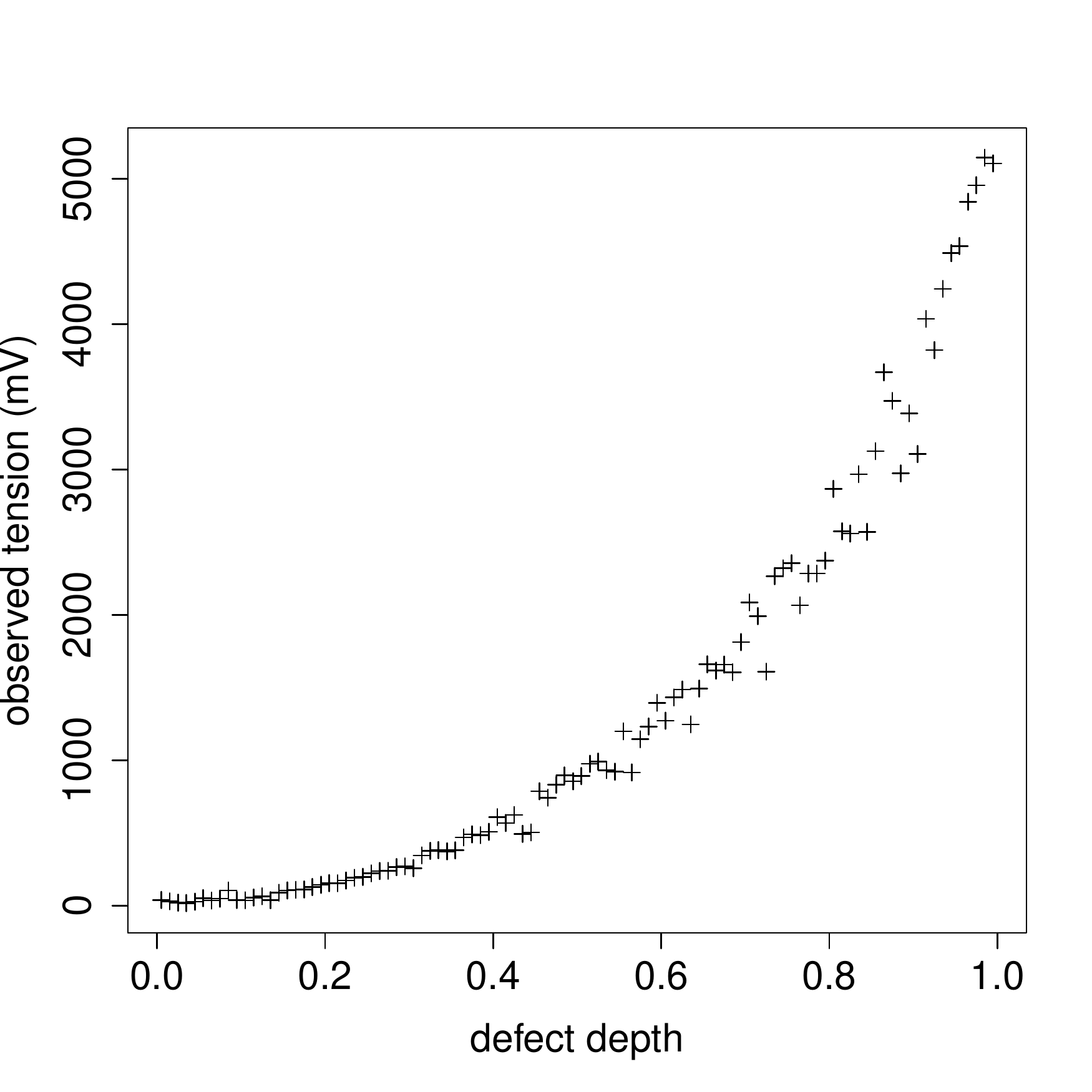}
\includegraphics[angle=0,width=0.33\linewidth, height=0.33\linewidth]{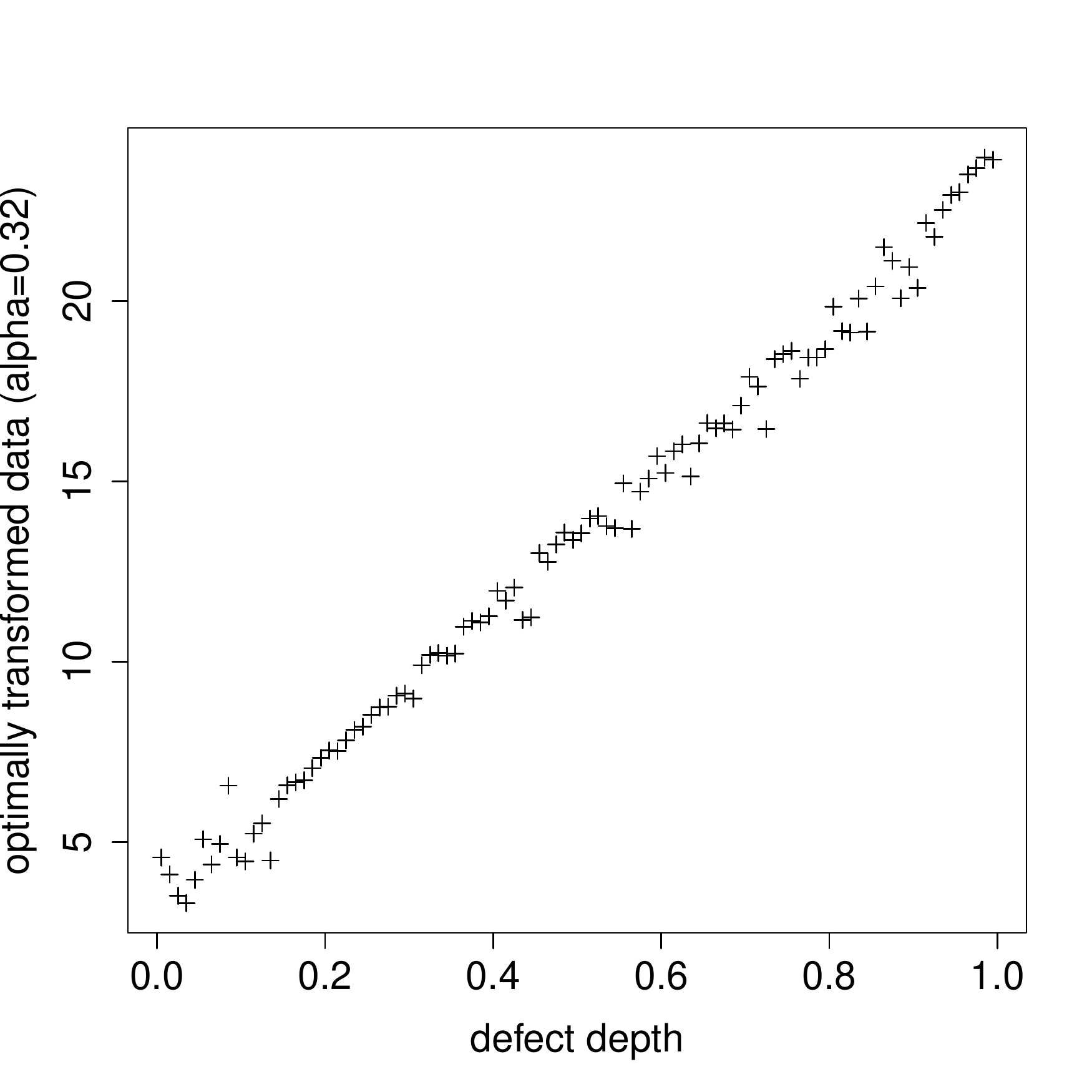}
\includegraphics[angle=0,width=0.33\linewidth, height=0.33\linewidth]{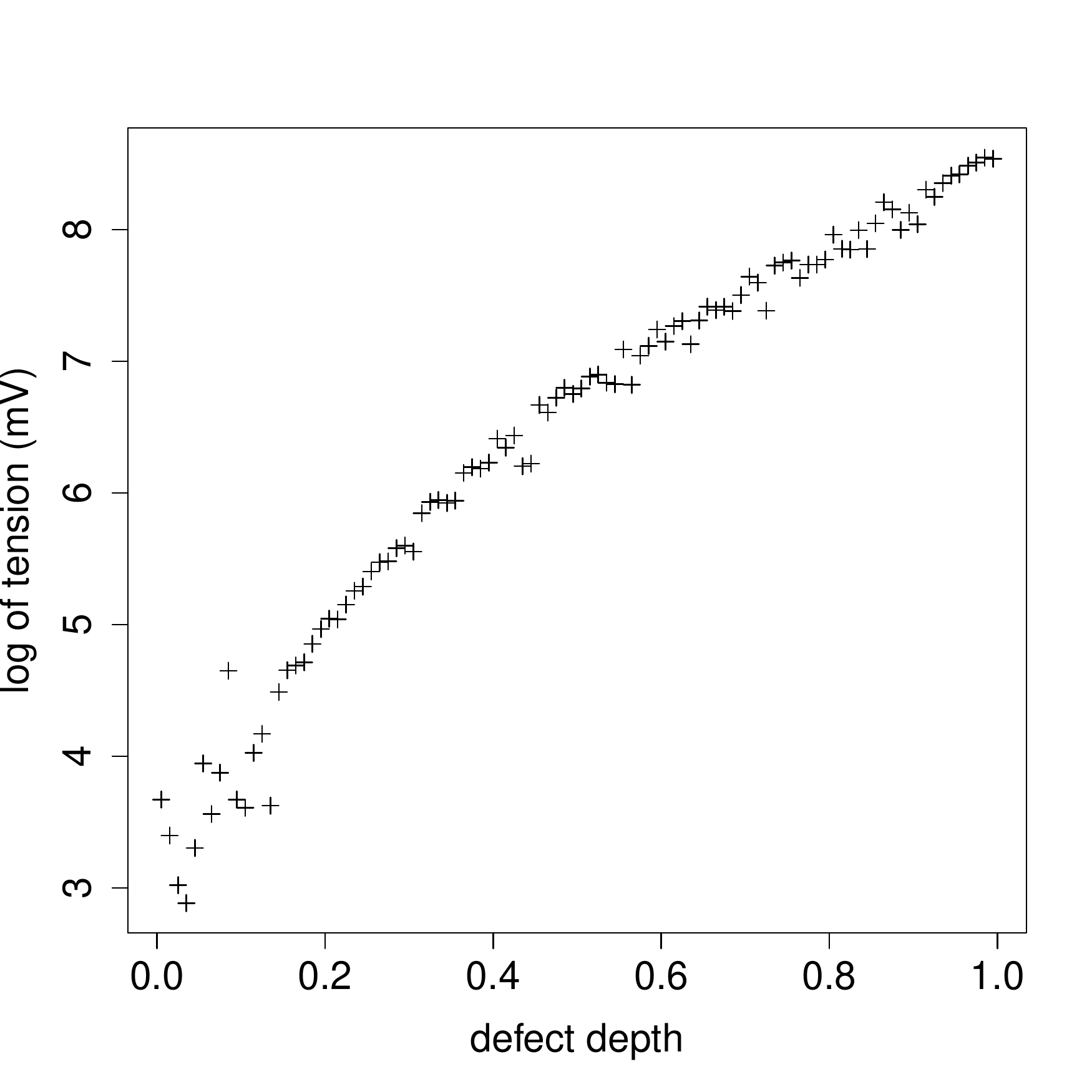}
\caption{Left: Measured tension as a mapping of defect depth for 100 defective tubes. The other ``nuisance'' geometrical parameters are representative possible geometries. Right: Same data after applying the logarithm function ($B_\alpha$ or $C_\alpha$ with $\alpha = 0$) to all measured tensions. Center: Same data after applying the optimal transformation ($C_\alpha$ with $\alpha=0.32$) to all measured tensions.}
\label{Fig:ObservedData}
\end{figure}

Figure \ref{Fig:ObservedData} (left) illustrates the importance of the length of the defect $a$. It represents the measured tensions for 100 defective tubes with various defect depths. The depths are normalized so that 1 represents the thickness of a tube coating. Using the parametrization presented at the end of the introduction, the 100 points form a space-filling design set for the 9-dimensional cube $(0,1)^9$. 

Because the measured tension is necessarily nonnegative, the Gaussian assumption of the Kriging surrogate model is inadequate. Moreover, the spread becomes greater and greater as the defect depth increases, which contradicts the assumption of stationarity.
Because Figure \ref{Fig:ObservedData} (left) gives the impression of being based on the graph of some exponential mapping, our first instinct was to apply the logarithm to the observations. The result is given by Figure \ref{Fig:ObservedData} (right).

Although this transformation seems fruitful, it is clearly too strong for our purpose. While the original data yielded what looked like a strongly convex mapping of defect depth, the log-transformed data yield a somewhat concave mapping of defect depth. Moreover, the spread which was originally increasing with depth seems now to be decreasing with depth. 

Some intermediate transformation between the identity mapping and the logarithm mapping was needed. A possible choice could have been the Box-Cox power transform family:

\begin{equation}
B_\alpha(t) = \left\{ \begin{array}{lr} \frac{t^{\alpha} - 1}{\alpha} & \text{if } \alpha > 0 \\ \log(t) & \text{if } \alpha = 0 \end{array} \right.
\end{equation}

The Box-Cox power transformation fits our requirements since all mappings $B_\alpha$ for $\alpha \in (0,1)$ are intermediate transformations between the logarithm ($\alpha=0$) and the identity mapping ($\alpha=1$, although the data are uniformly decremented by 1, which is of no consequence). However, for any $\alpha>0$, $B_\alpha$ is a bijection from $(0,+\infty)$ to $(-1/ \alpha, +\infty)$, whereas we would like a bijection from $(0,+\infty)$ to $(-\infty,+\infty)$, because otherwise how can the Gaussian assumption be credible ? In the Box-Cox family, only the logarithm mapping ($\alpha=0$) fits this requirement. \medskip

The following alteration to the Box-Cox family is suitable:

\begin{equation}
C_\alpha(t) = \left\{ \begin{array}{lr} \frac{1}{\alpha} \sinh(\alpha \log(t)) & \text{if } \alpha > 0 \\ \log(t) & \text{if } \alpha = 0 \end{array} \right.
\end{equation}

\begin{figure}[!ht]
\includegraphics[angle=0,width=0.5\linewidth, height=0.5\linewidth]{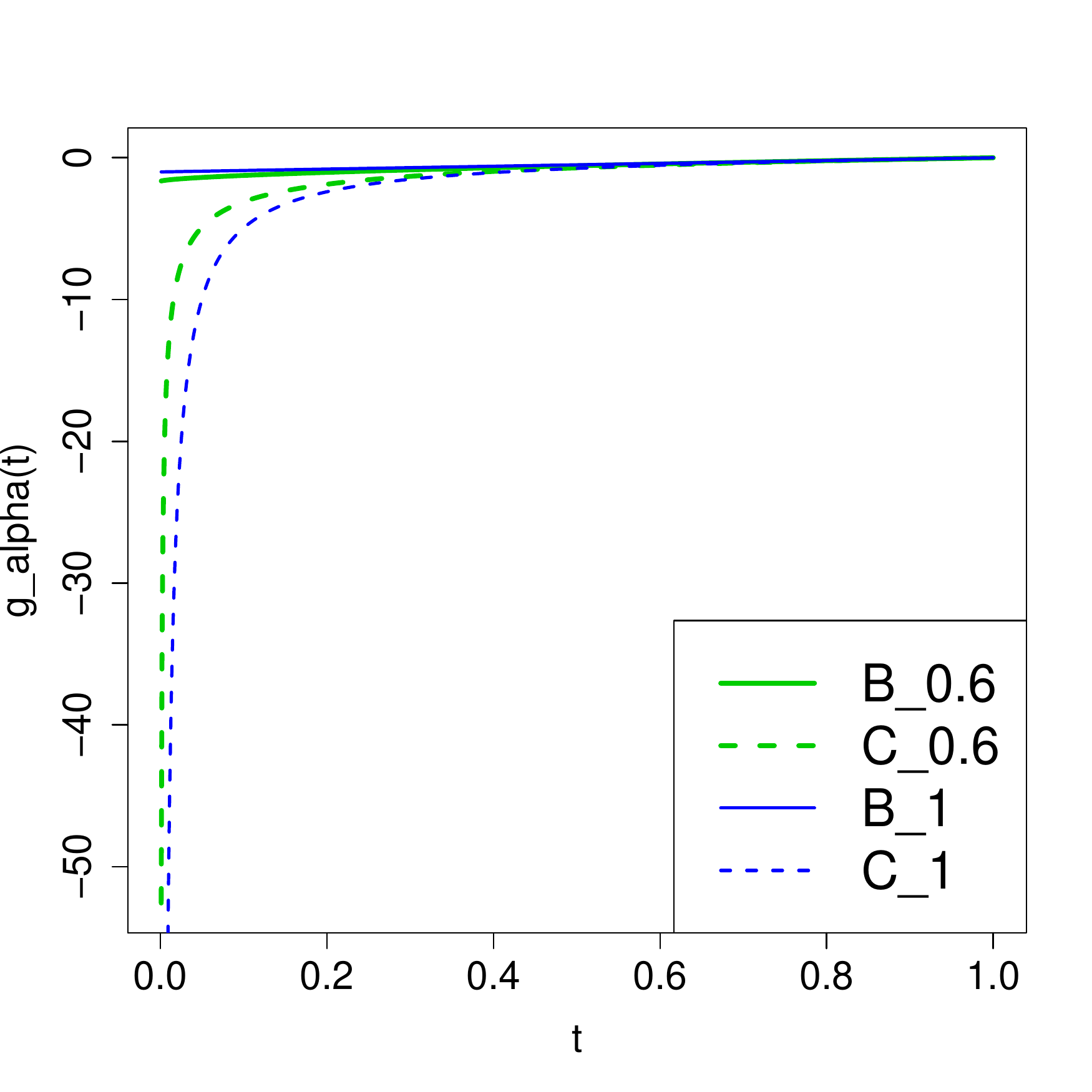}
\includegraphics[angle=0,width=0.5\linewidth, height=0.5\linewidth]{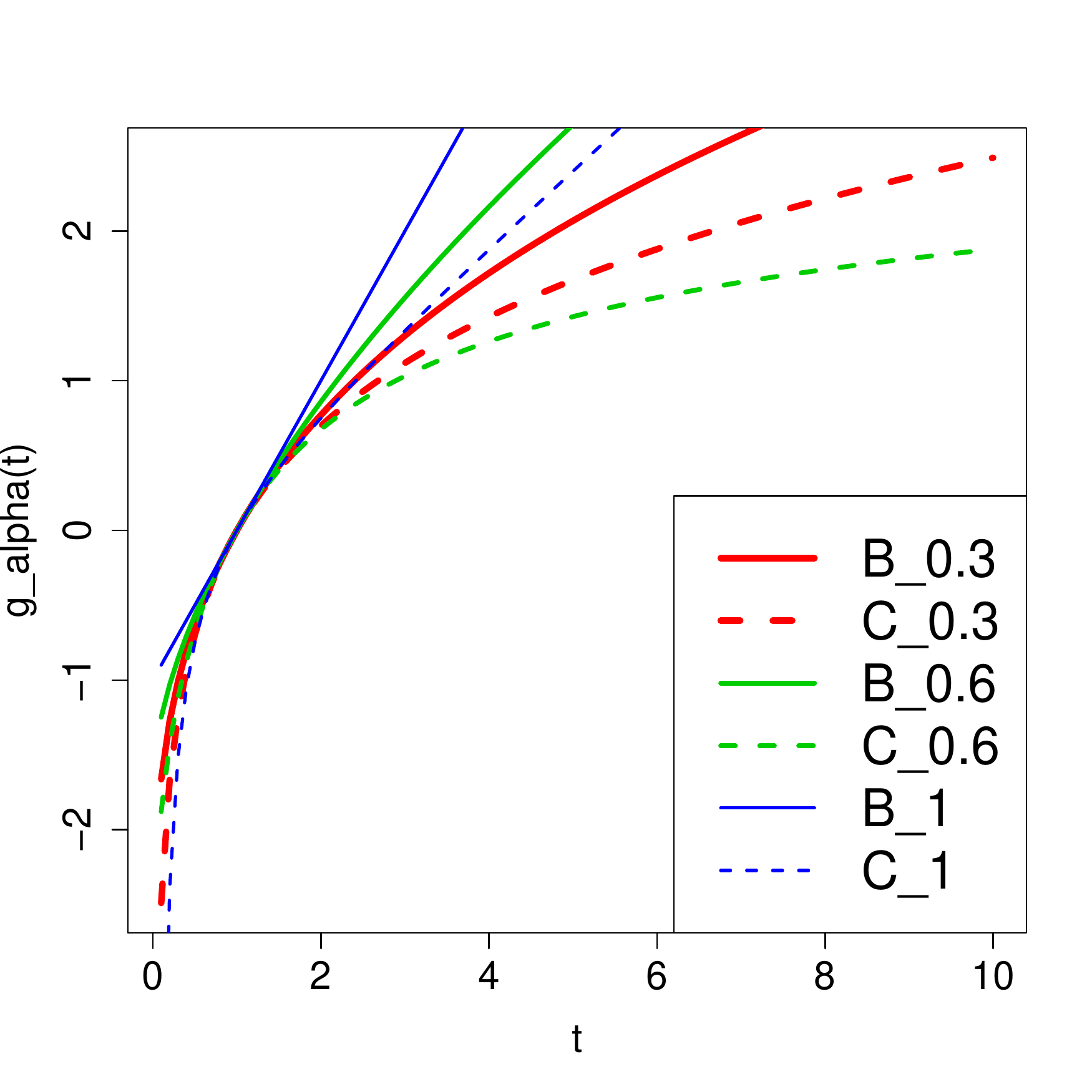}
\caption{Box-Cox and alternative transformation families for small values of $\alpha$ (left) and greater values of $\alpha$ (right). }
\label{Fig:transformation_families}
\end{figure}

Every mapping $C_\alpha$ is a bijection from $(0,+\infty)$ to $(-\infty,+\infty)$. Moreover, $C_1$ is equivalent to the linear mapping $t \mapsto 0.5 t$ when $t \to \infty$, so this family too can be considered to contain intermediate mappings between the logarithm ($\alpha=0$) and the identity mapping $(\alpha=1)$. Figure \ref{Fig:transformation_families} illustrates both Box-Cox and alternative transformation families. \medskip

We use this transformation family (i.e. $g_\alpha = C_\alpha$ for all $\alpha \in [0,+\infty)$) to apply the Bayesian framework for Trans-Gaussian Kriging described in the previous section. Our mean function space $\mathcal{F}_2$ is the space of affine functions of the parameter of interest $a$. This choice reflects the fact $a$ is the main influence on the output of the computer code. Our basis functions are the constant function of value 1 and the coordinate function $(a,\bs{x}) \mapsto a$. Our correlation kernel is the Matérn anisotropic geometric kernel of smoothness $\nu=5/2$. The transformation family is $C_\alpha$, $\alpha \in [0,+\infty)$. \medskip

\subsection{Trans-Gaussian Kriging Likelihood}

In Trans-Gaussian Kriging, $Z$ is a random field whose distribution is that of $\mathcal{Z}$ conditioned by $\mathcal{Z}(a^{(i)},\bs{x}^{(i)}) = z(a^{(i)},\bs{x}^{(i)})$ for every $i \in [\![1,n]\!]$, and $\bs{z}$ is the vector $\left(z(a^{(1)},\bs{x}^{(1)}),...,z(a^{(n)},\bs{x}^{(n)})\right) \trans$.

Denoting by $g_\alpha(\bs{z})$ the vector $(g_\alpha(z_1),...,g_\alpha(z_n)) \trans$,
the likelihood of hyperparameters  
$\bs{\beta}$, $\sigma^2$, $\bs{\theta}$ and $\alpha$ is:

\begin{equation} \label{Eq:vraisemblance_transgaussienne}
L( \bs{z} \; | \;  \bs{\beta}, \sigma^2 , \bs{ \theta }, \alpha ) = 
\left( \frac{ 1 }{ 2 \pi \sigma^2 } \right) ^ { \frac{n}{2} } | \bs{ \Sigma }_{ \bs{ \theta } } | ^ {- \frac{1}{2} } \exp \left\{ - \frac{ 1 }{ 2 \sigma^2 } \left(g_\alpha(\bs{z}) - \bs{H \beta} \right) \trans \bs{ \Sigma }_{ \bs{ \theta } }^{-1} \left( g_\alpha(\bs{z}) - \bs{H \beta} \right) \right\} 
\prod_{i=1}^n g_\alpha' (z_i) .
\end{equation}

With Trans-Gaussian Kriging, the ``Gaussian'' parameters $\bs{\beta}$, $\sigma^2$ and $\bs{\theta}$ can no longer be interpreted as pertaining to mean function, variance and correlation respectively. They can only be interpreted in relation to the transformation parameter $\alpha$.

\citep{BC64} propose estimating the ``Gaussian'' parameters conditionally to the transformation parameter, and to treat the value of the transformation parameter with maximum likelihood as the true value. This point of view was criticized by \citet{BD81}, who showed that when using the Box-Cox transformation family, the estimators for the transformation parameter and for $\bs{\beta}$ are highly correlated, and they argued for taking this effect into account when performing inference on the parameters.

Unfortunately, when the computational budget is too low to be able to make many observations, the Kriging likelihood can be rather flat even without adding a transformation parameter into the mix \citep{LS05}. \medskip

These difficulties are the reason why we propose in this paper using an Objective Bayesian framework to integrate $\bs{\beta}$, $\sigma^2$ and $\bs{\theta}$ out of the model -- conditionally to $\alpha$, in keeping with \citet{BC64}'s insight. This way, we only need to estimate the parameter $\alpha$, with the likely consequence that the ``integrated'' one-dimensional likelihood is not as flat as the regular multi-dimensional likelihood.

\subsection{Objective Bayesian treatment of the Gaussian parameters}

In this subsection, we operate again under the framework of Subsection \ref{Sec:Kriging_likeilhood}. 
This framework is the standard Kriging setting, which is parametrized by $\bs{\beta}$, $\sigma^2$ and $\bs{\theta}$. The likelihood is given by Equation (\ref{Eq:vraisemblance}).

Following \citet{BDOS01}, we make use of the reference prior on the parameters $(\bs{\beta},\sigma^2)$ conditional to $\bs{\theta}$: $\pi_R(\bs{\beta},\sigma^2 | \bs{\theta}) \propto 1 / \sigma^2$. The integrated likelihood $L^1(\bs{y} | \bs{\theta})$ is given by successively integrating $\bs{\beta}$ and $\sigma^2$ out. To do this, we introduce $\bs{W}$ as an $n \times (n-p)$ matrix such that $\bs{W} \trans \bs{H}$ is the null $(n-p) \times p$ matrix and $\bs{W} \trans \bs{W}$ is the $(n-p) \times (n-p)$ identity matrix. 
$\bs{W}$ can be computed by performing Standard Value Decomposition (SVD) on $\bs{H}$. The singular vectors corresponding to the null singular values form the columns of $\bs{W}$. We find

\begin{align}
L^1(\bs{y} | \bs{\theta}) = \int L(\bs{y} | \bs{\beta}, \sigma^2, \bs{\theta}) / \sigma^2 d \bs{\beta} d \sigma^2
 &= \left( \frac{2 \pi^{\frac{n-p}{2}}} {\Gamma\left( \frac{n-p}{2} \right)} \right)^{-1}  
 | \bs{W} \trans \bs{ \Sigma }_{ \bs{ \theta } } \bs{W} | ^ {- \frac{1}{2} }
  \left( \bs{y} \trans \bs{W} \left( \bs{W} \trans \bs{ \Sigma }_{ \bs{ \theta } } \bs{W} \right)^{-1} \bs{W} \trans \bs{y} \right)^{-\frac{n-p}{2} }. \label{Eq:vraisemblance_integree}
\end{align}

\newcommand{\HNouveauNouveau}{\bs{H}_{0,0}}
\newcommand{\HNouveauAncien}{\bs{H}_{0,\cdot}}
\newcommand{\HAncienNouveau}{\bs{H}_{\cdot,0}}
\newcommand{\corrNouveauNouveau}{\bs{\Sigma}_{\bs{\theta},0,0}}
\newcommand{\corrNouveauAncien}{\bs{\Sigma}_{\bs{\theta},0,\cdot}}
\newcommand{\corrAncienNouveau}{\bs{\Sigma}_{\bs{\theta},\cdot,0}}

\newcommand{\Emarg}{E_{0}}
\newcommand{\EmargW}{\bs{E}_{\bs{\theta},0}^{\bs{W}}}
\newcommand{\Vmarg}{S_{\bs{\theta},0,0}}
\newcommand{\VmargW}{\bs{S}_{\bs{\theta},0,0}^{\bs{W}}}
\newcommand{\corrmargAncienNouveau}{\bs{S}_{\bs{\theta},\cdot,0}}
\newcommand{\corrmargNouveauAncien}{\bs{S}_{\bs{\theta},0,\cdot}}
\newcommand{\corrmargWNouveauAncien}{\bs{S}_{\bs{\theta},0,\bs{W}}}
\newcommand{\corrmargWAncienNouveau}{\bs{S}_{\bs{\theta},\bs{W},0}}

We briefly explain how prediction can be performed in our Bayesian Kriging model. 

First, assume that all hyperparameters -- $\bs{\beta}$, $\sigma^2$ and $\bs{\theta}$ -- are known. We wish to predict the value taken by the Gaussian process $Y$ at an unobserved point $ (a_0,\bs{x}_0) $. \medskip 

Let $\bs{P}$ be the $n \times p$ matrix whose columns are obtained by applying the Gram-Schmidt process to the columns of $\bs{H}$. \medskip

Denote by $\HNouveauNouveau$ the transpose of the vector of length $p$ containing the values of the $p$  basis functions at$ (a_0,\bs{x}_0) $, by $\corrNouveauAncien$ the $1 \times n$ correlation matrix between the Gaussian process at $ (a_0,\bs{x}_0) $ and the Gaussian process at the design set, and finally by $\corrAncienNouveau$ its transpose.

Define the $1 \times p$ matrix $\HNouveauAncien = \HNouveauNouveau \left( \bs{P} \trans \bs{H} \right)^{-1}$ and its transpose $\HAncienNouveau$. The following matrix definitions are necessary to express the predictive distribution:

\begin{align}
\Emarg :&= \HNouveauAncien \bs{P} \trans \bs{y} \nonumber \\ 
\Vmarg 
:&=
1 +
\HNouveauAncien \bs{P} \trans \corr \bs{P} \HAncienNouveau
- \HNouveauAncien \bs{P} \trans \corrAncienNouveau - \corrNouveauAncien \bs{P} \HAncienNouveau
\nonumber \\
\corrmargNouveauAncien  
:&=
\left(
\HNouveauAncien \bs{P} \trans \corr - \corrNouveauAncien
\right)
\left( \bs{P} \bs{W} \right)
\nonumber \\
\corrmargAncienNouveau :&= \corrmargNouveauAncien \trans
\nonumber 
\end{align}

The following result is proved in \citep{Mur18b}.

\begin{prop}
The predictive distribution of $Y(a_0,\bs{x}_0)$ averaged over both $\bs{\beta}$ and $\sigma^2$ is the non-standardized Student's t-distribution with $n-p$ degrees of freedom, location parameter
$\Emarg - \corrmargNouveauAncien \bs{W} \left( \bs{W} \trans \corr \bs{W} \right)^{-1} \bs{W} \trans \bs{y}$
and scale parameter
$$ \sqrt{ \frac{\bs{y} \trans \bs{W} \left( \bs{W} \trans \corr \bs{W} \right)^{-1} \bs{W} \trans \bs{y}}{n-p}\left\{ \Vmarg - \corrmargNouveauAncien \bs{W} \left( \bs{W} \trans \corr \bs{W} \right)^{-1} \bs{W} \trans \corrmargAncienNouveau \right\} }.$$
\end{prop}

Practically speaking, if $n-p$ is large the Student t-distribution can be approximated by a Normal distribution with mean equal to the location parameter and standard deviation equal to the scale parameter. \medskip

Further integrating the predictive distribution requires averaging over the Gibbs reference posterior distribution on $\bs{\theta}$. The Gibbs reference posterior distribution is derived through Objective Bayesian techniques and has nice theoretical as well as practical properties, like invariance by reparametrization and good frequentist performance for prediction intervals \citep{Mur18b}. This integration can be done numerically, by using a sample from the Gibbs reference posterior that can be obtained easily by a Gibbs sampler. \medskip

\subsection{The Likelihood problem}

Let us go back to Trans-Gaussian Kriging. In this framework, it is desirable to integrate $\bs{\theta}$ out of the model in order to derive an integrated likelihood of the transformation parameter $\alpha$ only.

The problem with the Gibbs reference posterior approach is that, given there is no actual prior distribution yielding the Gibbs reference posterior, it is not possible to integrate $\bs{\theta}$ out of the model with likelihood $L^1(\bs{y} | \bs{\theta})$. This is a major hurdle since the actual model (taking into account that $\bs{y}$ is the result of a transformation) is:

\begin{equation}
L^1(\bs{z} | \bs{\theta}, \alpha) =
 \left( \frac{2 \pi^{\frac{n-p}{2}}} {\Gamma\left( \frac{n-p}{2} \right)} \right)^{-1}  
 | \bs{W} \trans \bs{ \Sigma }_{ \bs{ \theta } } \bs{W} | ^ {- \frac{1}{2} }
  \left( g_\alpha(\bs{z}) \trans \bs{W} \left( \bs{W} \trans \bs{ \Sigma }_{ \bs{ \theta } } \bs{W} \right)^{-1} \bs{W} \trans g_\alpha(\bs{z}) \right)^{-\frac{n-p}{2} } 
  \prod_{i=1}^n g_\alpha' (z_i) . \label{Eq:vraisemblance_integree_transformation}
\end{equation}

Because we cannot integrate $\bs{\theta}$ out, we need to compute some other aggregate of all possible $L^1(\bs{z} | \bs{\theta}, \alpha)$ when $\bs{\theta}$ varies. The most obvious solution is to average $L^1(\bs{z} | \bs{\theta}, \alpha)$ over the Gibbs reference posterior distribution $\pi_G(\bs{\theta} | \bs{z}, \alpha) := \pi_G(\bs{\theta} | g_\alpha(\bs{z}))$, but this solution gives too much weight to the likelihood: if a prior $\pi_G(\bs{\theta}|\alpha)$ existed, then this would be equivalent to computing $\int L^1(\bs{z} | \bs{\theta}, \alpha)^2 \pi_G(\bs{\theta}|\alpha) d\bs{\theta} / \int L^1(\bs{z} | \bs{\theta}, \alpha) \pi_G(\bs{\theta}|\alpha) d\bs{\theta}$. \medskip

A second possibility is to compute the Maximum A Posteriori estimator $\widehat{\bs{\theta}}_{MAP}$ for $\bs{\theta}$ and then proceed with $L^{MAP}(\bs{z} | \alpha) := L^1(\bs{z} | \bs{\theta} = \widehat{\bs{\theta}}_{MAP}, \alpha)$. The idea is that $\widehat{\bs{\theta}}_{MAP}$ should be a ``typical'' value of $\bs{\theta}$, but this approach unfortunately makes the aggregate dependent on the parametrization chosen for the correlation lengths, which is a bad property for a quantity treated as a likelihood. \medskip

A third possibility is 

\begin{equation}
L^{LOG}(\bs{z} | \alpha) := \exp \left[ \int \log \{L^1(\bs{z} | \bs{\theta}, \alpha)\} \pi_G(\bs{\theta}|\bs{z}, \alpha) d\bs{\theta} \right]
\end{equation}

Compared to $L^{MAP}$, $L^{LOG}$ has the advantage that it does not depend on the parametrization, insofar as $\bs{\theta}$ could be replaced by any vector $(h_1(\theta_1),...,h_r(\theta_r)) \trans$ as long as all $h_i$ ($i \in [\![1,r]\!]$) are bijections. Moreover, it does not rely on a particular estimate of the ``true'' $\bs{\theta}$, which makes $L^{LOG}$ more robust that $L^{MAP}$. \medskip

Fundamentally though, $L^{MAP}$ and $L^{LOG}$ are justified by a simple heuristic. This heuristic is based on two 
approximations. \medskip

\begin{enumerate}
\item Asymptotic: the Gibbs reference posterior distribution is approximated by $\mathcal{N}(\widehat{\bs{\theta}}_{MAP} ; \mathcal{I}( \widehat{\bs{\theta}}_{MAP} )^{-1} )$ where $\mathcal{I}( \bs{\theta} )$ is the Fisher information matrix. This means that we assume that the conclusion of the Bernstein - von Misès theorem applies as though we were in an asymptotic framework.
\item Jeffreys: the Gibbs reference posterior is the posterior distribution of $\bs{theta}$ corresponding to the Jeffreys-rule prior on $\bs{\theta}$ denoted by $\pi(\bs{\theta} | \alpha)$. In order to have simple notations, we do not normalize it -- it is possibly not proper anyway -- so, denoting by $| \cdot |$ the matrix determinant, $\pi(\bs{\theta} | \alpha) = |  \mathcal{I}(\bs{\theta} ) |^{1/2}$.
\end{enumerate}

Define $K_{\bs{z}}(\alpha) \in \R$ such that $ K_{\bs{z}}(\alpha) \int L^1(\bs{z} | \bs{\theta}, \alpha) \pi(\bs{\theta} | \alpha) d \bs{\theta} = \prod_{i=1}^n g_\alpha' (z_i)$. This is equivalent to asserting that $\pi_G(\bs{\theta} | \bs{z}, \alpha) \prod_{i=1}^n g_\alpha' (z_i) = K_{\bs{z}}(\alpha)  L^1(\bs{z} | \bs{\theta}, \alpha) \pi(\bs{\theta} | \alpha)$. In the following, we show that the two approximations imply that $\prod_{i=1}^n g_\alpha' (z_i) / K_{\bs{z}}(\alpha)$ is related to both $L^{MAP}(\bs{z} | \alpha)$ and $L^{LOG}(\bs{z} | \alpha)$. \medskip

First, we have

\begin{equation}
K_{\bs{z}}(\alpha) L^1(\bs{z} | \bs{\theta}, \alpha) \pi(\bs{\theta} | \alpha) = (2\pi)^{-r/2} | \mathcal{I} (\bs{\theta}) |^{1/2} 
\exp \left\{ - \frac{ (\bs{\theta} - \widehat{\bs{\theta}}_{MAP}) \trans \mathcal{I}( \widehat{\bs{\theta}}_{MAP} ) (\bs{\theta} - \widehat{\bs{\theta}}_{MAP}) }{2} \right\}
\prod_{i=1}^n g_\alpha' (z_i) .
\end{equation}

In particular,

\begin{equation}
K_{\bs{z}}(\alpha) L^1(\bs{z} | \widehat{\bs{\theta}}_{MAP}, \alpha)  = (2\pi)^{-r/2}  \prod_{i=1}^n g_\alpha' (z_i) .
\end{equation}

So the integrated likelihood $\int L^1(\bs{z} | \bs{\theta}, \alpha) \pi(\bs{\theta} | \alpha) d \bs{\theta}$ is proportional to $L^{MAP}(\bs{z} | \alpha)$:

\begin{equation}
\int L^1(\bs{z} | \bs{\theta}, \alpha) \pi(\bs{\theta} | \alpha) d \bs{\theta} = (2\pi)^{r/2}  L^{MAP}(\bs{z} | \alpha).
\end{equation}

Furthermore, 

\begin{equation}
\int \log L^1(\bs{z} | \bs{\theta}, \alpha) \pi_G(\bs{\theta} | \bs{z}, \alpha) d \bs{\theta}
=
\int \log \{ K_{\bs{z}}(\alpha) L^1(\bs{z} | \bs{\theta}, \alpha) \} \pi_G(\bs{\theta} | \bs{z}, \alpha) d \bs{\theta} - \log K_{\bs{z}}(\alpha)
\end{equation}

Notice that

\begin{equation}
\log \{ K_{\bs{z}}(\alpha) L^1(\bs{z} | \bs{\theta}, \alpha) \} = -\frac{r}{2} \log(2\pi) -  \frac{ (\bs{\theta} - \widehat{\bs{\theta}}_{MAP}) \trans \mathcal{I}( \widehat{\bs{\theta}}_{MAP} ) (\bs{\theta} - \widehat{\bs{\theta}}_{MAP}) }{2}
+ \sum_{i=1}^n \log g_\alpha'(z_i),
\end{equation}

so

\begin{equation}
\int \log L^1(\bs{z} | \bs{\theta}, \alpha) \pi_G(\bs{\theta} | \bs{z}, \alpha) d \bs{\theta}
=
-\frac{r}{2} \log(2\pi) - \frac{r}{2} + \sum_{i=1}^n \log g_\alpha'(z_i) - \log K_{\bs{z}}(\alpha).
\end{equation}

Finally, we see that the integrated likelihood $\int L^1(\bs{z} | \bs{\theta}, \alpha) \pi(\bs{\theta} | \alpha) d \bs{\theta}$ is also proportional to $L^{LOG}(\bs{z} | \alpha)$:

\begin{equation}
\log \left\{ \int L^1(\bs{z} | \bs{\theta}, \alpha) \pi(\bs{\theta} | \alpha) d \bs{\theta} \right\} = \log L^{LOG}(\bs{z}|\alpha) + \frac{r}{2} \log(2\pi) + \frac{r}{2}.
\end{equation}

Interestingly, this heuristic also predicts that 

\begin{equation} \label{Eq:prediction_heuristic}
\log \{ L^{MAP}(\bs{z} | \alpha) \} - \log \{ L^{LOG}(\bs{z} | \alpha) \} = r/2.
\end{equation}

This prediction provides a sanity check for the heuristic, which we use in the application below.

\section{Industrial Application} \label{Sec:Industrial_application}

\subsection{Integrating out Kriging hyperparameters}

Figure \ref{Fig:ObservedData} shows that $\alpha$ can only reasonably belong to $[0,1]$, so we endow $[0,1]$ with the fine grid $0.01 * [\![0,100]\!]$. For every element $\alpha$ in this grid, we sample 100 points $\bs{\theta}_\alpha^{(j)} \in (0,+\infty)^r$ from the Gibbs reference posterior distribution. This is done by performing 9000 iterations of the Gibbs algorithm and keeping only one out of 90.

\begin{figure}[!ht]
\includegraphics[angle=0,width=0.5\linewidth, height=0.5\linewidth]{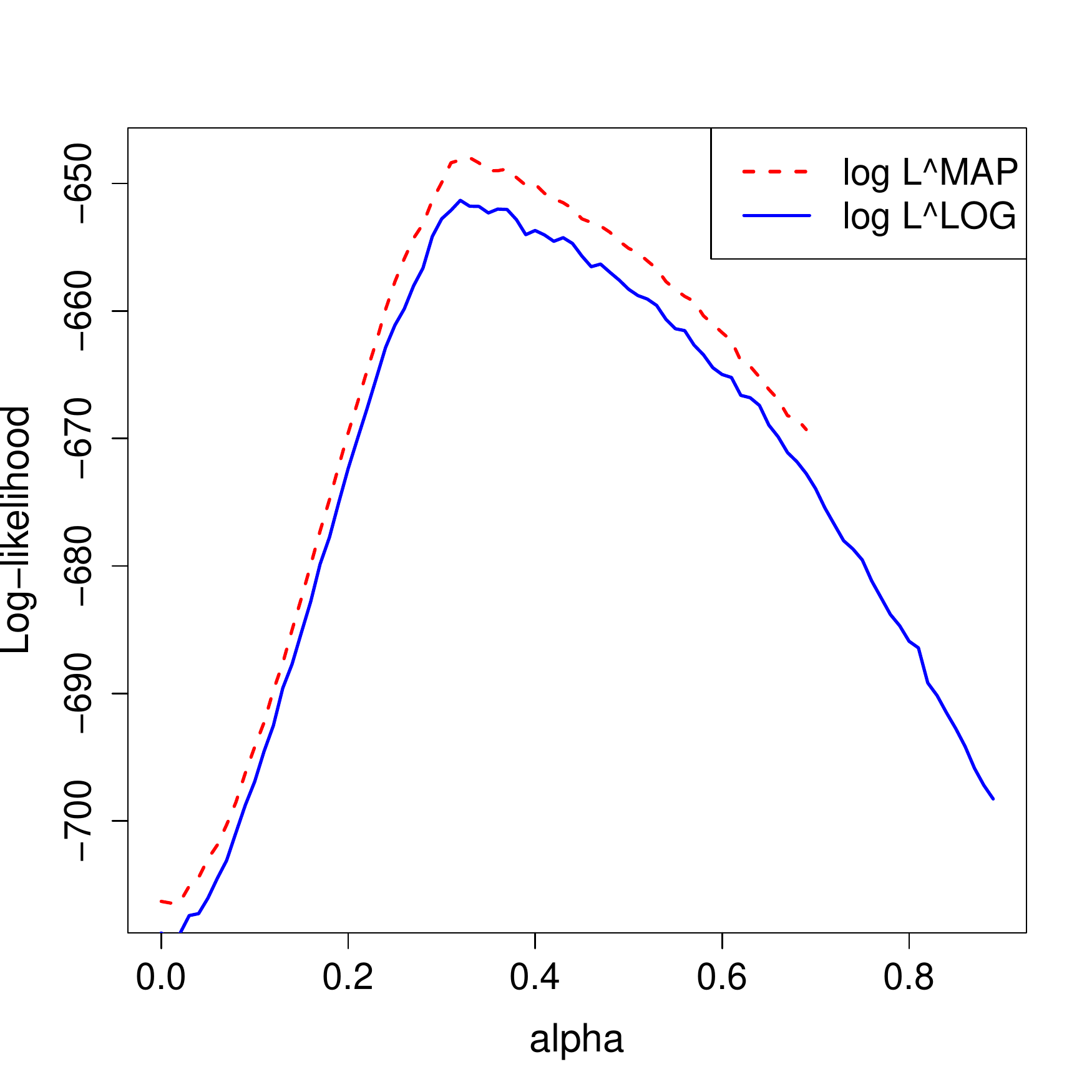}
\includegraphics[angle=0,width=0.5\linewidth, height=0.5\linewidth]{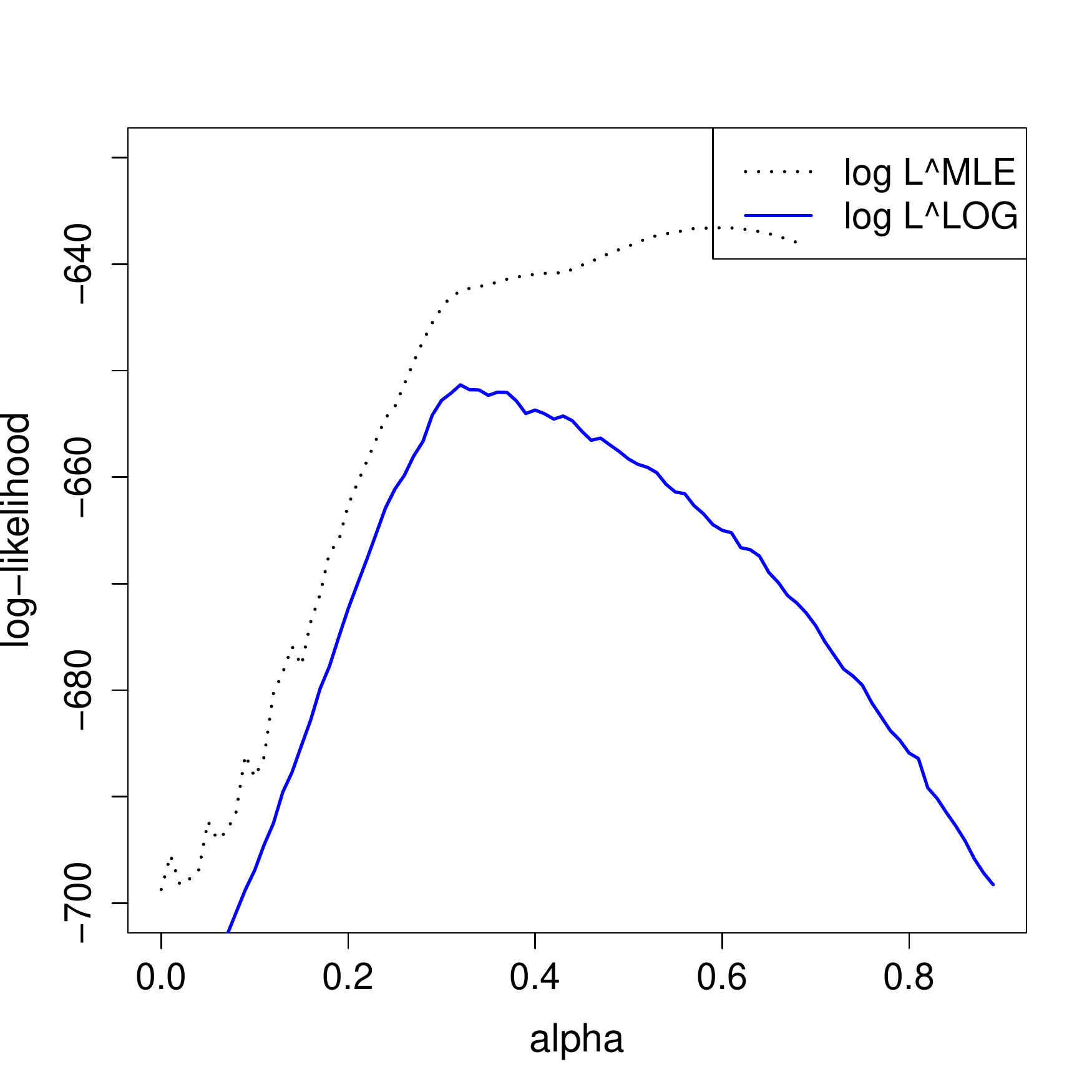}
\caption{Left: Logarithm of $L^{MAP}(\bs{z}|\alpha)$ (red dotted line) and $L^{LOG}(\bs{z}|\alpha)$ (blue solid line). $L^{MAP}$ is not represented for greater values of $\alpha$ because the MAP cannot be reliably computed. Right: Logarithm of $L^{LOG}(\bs{z}|\alpha)$ (blue solid line) 
and log-likelihood $L^{MLE}(\bs{z}|\alpha)$ if the MLE estimator for $\bs{\theta}$ is used (black dotted line). $L^{MAP}$ and $L^{MLE}$ are not represented for greater values of $\alpha$, because the MAP cannot be reliably computed and the $MLE$ favors correlation lengths so high that correlation matrices are ill-conditioned.}
\label{Fig:logLikelihood}
\end{figure}

Figure \ref{Fig:logLikelihood} (left) gives the logarithm of the pseudo-likelihoods $L^{MAP}$ and $L^{LOG}$.

$L^{MAP}$ and $L^{LOG}$ reach their maxima at $\alpha=0.34$ and $\alpha=0.32$ respectively. Interestingly, the average value of $L^{MAP}(\bs{z}|\alpha) - L^{LOG}(\bs{z}|\alpha)$ for $\alpha \in [0.25,0.55]$ 
is 3.4 and its standard deviation 0.3. 
This is quite close to the prediction (\ref{Eq:prediction_heuristic}) that this difference should be constant, especially considering that the values of both functions spread over an interval of length greater than 50. However, the value 3.4 is substantially lower than the predicted $r/2=4.5$, which suggests that not all 9 parameters are influent. \medskip

On account of $L^{LOG}$ not depending on the estimate of the MAP, we accept the value of $\alpha$ maximizing  $L^{LOG}$, $\alpha=0.32$ as the ``true'' alpha. In this respect, we follow the suggestion of \citet{BC64}. Naturally, a completely Bayesian treatment would require defining a prior distribution on $\alpha$ in order to integrate it out. In this case however, both versions of the likelihood are so pronounced that any reasonable prior would have little effect on the posterior distribution. In the interest of completeness though, we do offer in the next section a comparison between the prediction of the Maximum Likelihood approach setting $\alpha=0.32$ and a ``full-Bayesian'' approach using the Uniform prior distribution on $[0,1]$. \medskip

The question of whether or not to use a Bayesian approach to deal with $\alpha$ instead of the MLE may be debatable, but it is not when dealing with $\bs{\theta}$: the Bayesian approach is clearly superior. Figure \ref{Fig:logLikelihood} (right) shows the log-likelihood of $\alpha$ when taking $\bs{\theta}$ to be equal to its MLE (dotted curve). It favors high values of $\alpha$, which means the data are weakly transformed so the MLE on $\bs{\theta}$ favors very high correlation lengths and correlation matrices become ill-conditioned.  

Let us consider again the observation data presented in Figure \ref{Fig:ObservedData} (left). If we apply the transformation $C_\alpha$ with $\alpha=0.32$, we obtain Figure \ref{Fig:ObservedData} (center). Notice that the observations seem to be placed along a straight line. This is no coincidence: it reflects our choice of mean function space $\mathcal{F}_2$. The optimal transformation parameter is in first approximation the one that makes the data match our assumption about the mean function.

To confirm this, consider a surrogate model of transformed linear regression. It can be seen as a Trans-Gaussian Kriging model with null correlation lengths. We provide the log-likelihood of the transformation parameter for such a model in Figure \ref{Fig:logLikelihood} (right). It reaches its maximum at $\alpha=0.30$, very near $\alpha=0.32$ where $L^{LOG}$ reaches its maximum. This shows that the correlation structure parametrized by $\bs{\theta}$ has little impact on the maximum likelihood of $\alpha$. $\bs{\beta}$ and $\sigma^2$ having been integrated out of the model, the choice of mean function space $\mathcal{F}_p$ is necessarily the primary explanation for the the likelihood function of $\alpha$ reaching its maximum where it does.

\subsection{Probability Of defect Detection}

With this machinery, we can now return to the question of POD curves. For any point in the $r=9$-dimensional input space of C3D, the Trans-Gaussian Kriging surrogate model can provide a probability that the detection threshold $s=200mV$ is crossed. 

We need a few definitions

\begin{defn}
Let $a \in [0,1]$. The $\emph{actual POD}$, denoted by $POD(a)$, is the probability for a defect of depth $a$ to be detected: $POD(a) = \mathbb{P}(z(a,\bs{X})>s)$
\end{defn}

This probability refers to actual randomness, in the sense that the geometrical characteristics of a defect are considered random. It is a probability in the frequentist sense but cannot be accessed without prohibitive computational costs because it would involve running C3D over a large set of 
geometries $\bs{x}$.

\begin{defn}
The \emph{surrogate model safety} denoted by $SAFE(a,\bs{x})$ is the probability, according to the surrogate model, of a particular defect characterized by $(a,\bs{x})$ being detected:   $SAFE(a,\bs{x}) = \mathbb{P}(Z(a,\bs{x})>s)$.
\end{defn}

Contrarily to actual POD, surrogate model safety does not refer to any randomness in the frequentist meaning but instead expresses the ``opinion'' of the surrogate model. It represents epistemic uncertainty and is a probability in the Bayesian sense.

\begin{defn}
Let $a \in [0,1]$. The \emph{mean POD}, denoted by $POD_{mean}(a)$, is the average of surrogate model safety over all defects of depth $a$: $POD_{mean}(a) = \mathbb{E}(SAFE(a,\bs{X})) = \mathbb{P}(Z(a,\bs{X})>s)$.
\end{defn}

The mean POD is perhaps the best approximation of the actual POD available to us. However it is difficult to interpret since it aggregates two very different kinds of uncertainty: uncertainty about the defect geometry, which is random in kind, and epistemic uncertainty, which refers to imprecision on the part of the surrogate model.

\begin{defn}
Let $a \in [0,1]$ and $\gamma \in [0,1]$. The \emph{POD} at safety level $\gamma$, denoted by $POD_\gamma(a)$, is the probability for surrogate model safety to be greater or equal to $\gamma$: 

$POD_\gamma(a) = \mathbb{P}(SAFE(a,\bs{X}) \geqslant \gamma)$.

\end{defn}

The POD at safety level $\gamma$ for a given depth length $a$ is the probability of the surrogate model being confident about the defect being detected, with $\gamma$ denoting the required confidence level. Its aim is to constrain epistemic uncertainty in order to provide a figure that reflects actual randomness. Its interpretation is therefore clearer than that of the mean POD. However, it is of interest only if $\gamma$ is high. In the following, we compute it with $\gamma=95\%$ and $\gamma=99\%$. \medskip

Computationally speaking, because the defect depth $a$ belongs to the interval $[0,1]$, we endow this interval with the fine grid $0.01 * [\![0,100]\!]$. For every value of $a$ in this grid, we generate a 1000-points sample from the probability distribution of $\bs{X}$. For every $a$ therefore, we gather 1000 probabilities of defect detection. With this, we may:

\begin{enumerate}
\item compute the mean;

\item count how many are greater or equal to safety levels (95\% and 99\%).
\end{enumerate}

The first quantity is a Monte-Carlo approximation of the mean POD, the second of the POD at safety level 95\% or 99\%. \medskip

\begin{figure}[!ht]
\includegraphics[angle=0,width=0.5\linewidth, height=0.5\linewidth]{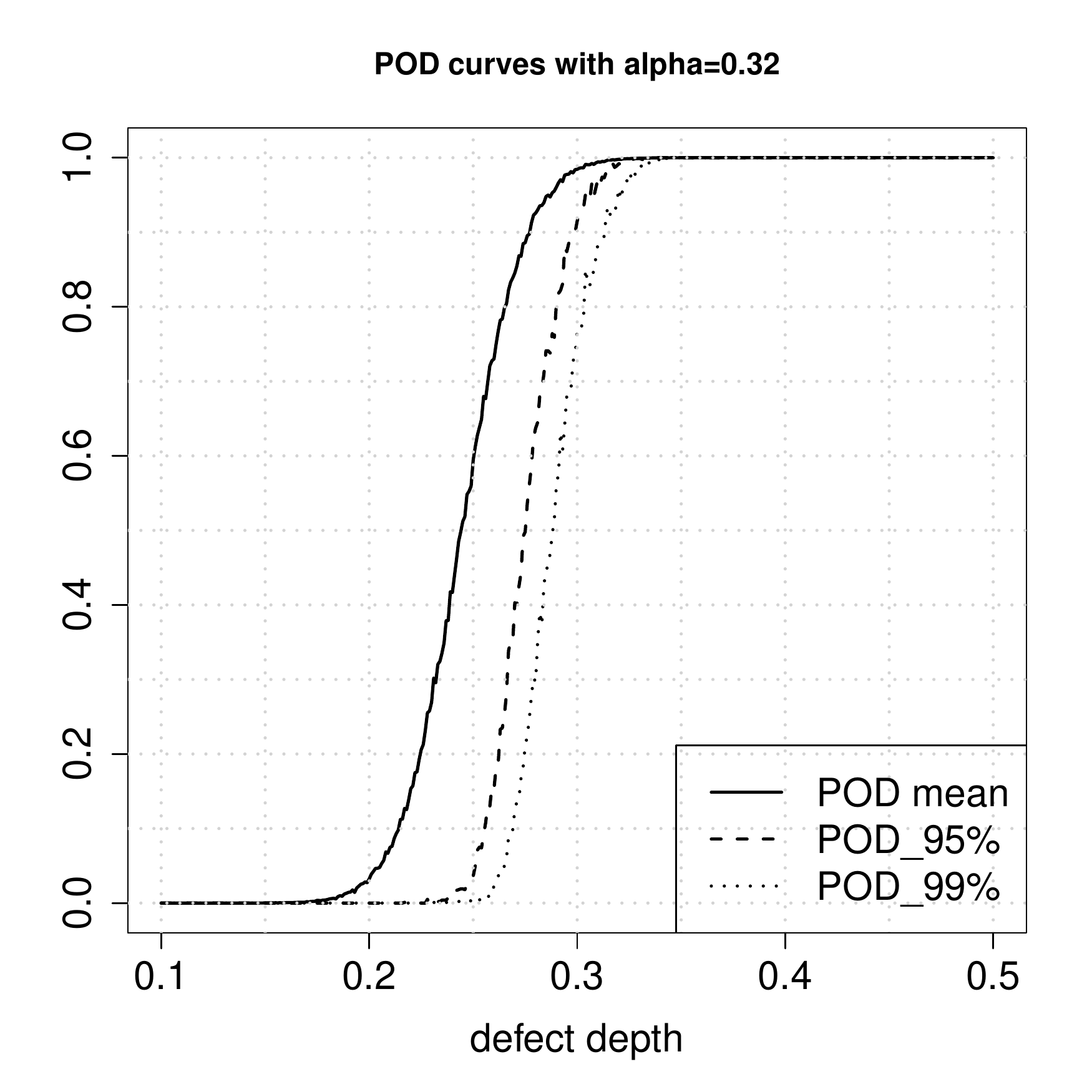}
\includegraphics[angle=0,width=0.5\linewidth, height=0.5\linewidth]{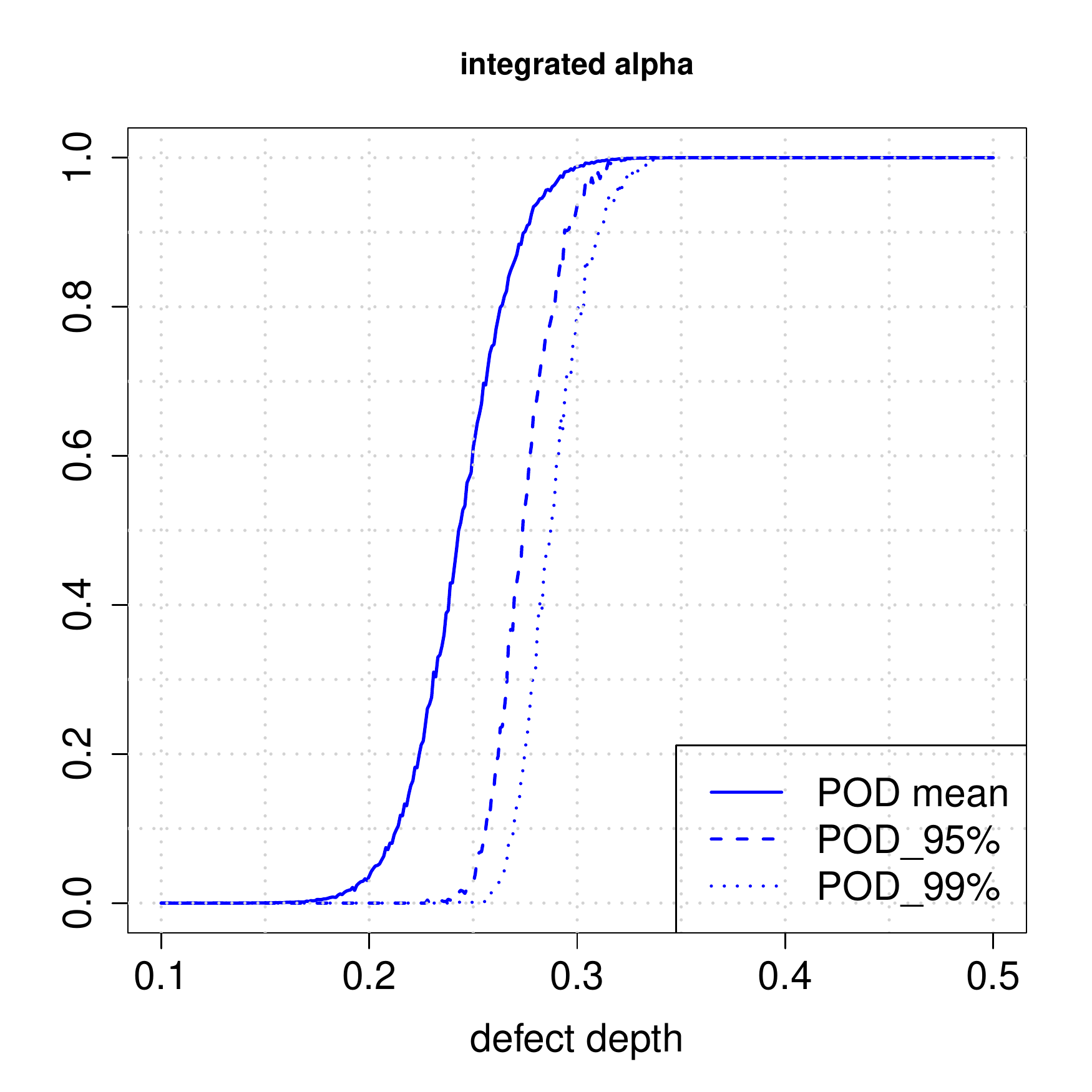}
\caption{Left: POD curves with $\alpha=0.32$. Right: POD curves with $\alpha=0.32$ and for integrated $\alpha$ (points). Since curves with $\alpha=0.32$ and integrated $\alpha$ are almost confused, it was not necessary to use different graphical styles for the different POD curves derived with integrated $\alpha$.}
\label{Fig:POD}
\end{figure}

The mean POD and POD at safety level 95\% and 99\% are drawn in Figure \ref{Fig:POD}. On the left, $\alpha$ is taken to be 0.32, the value for which $L^{LOG}$ reaches its maximum. On the right, they are also depicted and accompanied by the curves obtained with integrated $\alpha$. \medskip

Integrated $\alpha$ here means that the Uniform prior distribution on $[0,1]$ was used for transformation parameter $\alpha$. Because the likelihood $L^{LOG}$ has a very sharp peak, integration was performed only over the interval $[0.30,0.37]$. The complement set was deemed to have too low posterior probability to be worth taking into account. Integration was performed by using the rectangle method over $[0.30,0.37]$ with gap $0.01$. This is admittedly a rough approximation, and is intended to serve as a glimpse into a Bayesian treatment of every single parameter of the problem. Figure \ref{Fig:POD} (right) shows that the full-Bayesian method differs only slightly from the Maximum Likelihood method. Interestingly, the full-Bayesian method seems slightly less conservative than the Maximum Likelihood method. We therefore adopt in this case the Maximum Likelihood method for simplicity and conservativeness both.

\section{Conclusion}

In this work, we demonstrated how an Objective Bayesian framework could be applied to a Trans-Gaussian Kriging surrogate model. Contrary to Maximum Likelihood approaches, it makes the likelihood function of the transformation parameter clearly discriminate between all possible candidates. Moreover, it provides a way to naturally incorporate hyperparameter uncertainty into the prediction (at least as far as ``Gaussian parameters'' are concerned, because one of our findings was precisely that there is not much uncertainty about the transformation parameter). \medskip 

This is especially useful in the context of calibration of NDT techniques by numerical simulation. It makes a clear distinction between randomness and epistemic uncertainty possible. This distinction can then be incorporated into Probability Of defect Detection (POD) curves. \medskip

Concerning the particular problem of detecting defects in Steam Generator tubes, our framework provides a theoretically sound solution for estimating surrogate model uncertainty. This makes interpretation of results easier and thereby increases the reliability of the results of the study.

\section*{Acknowledgements}
The author would like to thank his PhD advisor Professor Josselin Garnier (École Polytechnique, Centre de Mathématiques Appliquées) for his guidance, Loic Le Gratiet (EDF R\&D, Chatou) and Anne Dutfoy (EDF R\&D, Saclay) for their advice and helpful suggestions.
The author acknowledges the support of the French Agence Nationale de la Recherche (ANR), under grant ANR-13-MONU-0005 (project CHORUS).

\pagebreak
\bibliography{biblio}
\end{document}